# The Availability and Persistence of Web References in D-Lib Magazine

Frank McCown, Sheffan Chan, Michael L. Nelson, Johan Bollen

Old Dominion University
Department of Computer Science
Norfolk, VA 23529 USA
{fmccown,chan_s,mln,jbollen}@cs.odu.edu

**Abstract.** We explore the availability and persistence of URLs cited in articles published in D-Lib Magazine. We extracted 4387 unique URLs referenced in 453 articles published from July 1995 to August 2004. The availability was checked three times a week for 25 weeks from September 2004 to February 2005. We found that approximately 28% of those URLs failed to resolve initially, and 30% failed to resolve at the last check. A majority of the unresolved URLs were due to 404 (page not found) and 500 (internal server error) errors. The content pointed to by the URLs was relatively stable; only 16% of the content registered more than a 1 KB change during the testing period. We explore possible factors which may cause a URL to fail by examining its age, path depth, top-level domain and file extension. Based on the data collected, we found the half-life of a URL referenced in a D-Lib Magazine article is approximately 10 years. We also found that URLs were more likely to be unavailable if they pointed to resources in the .net, .edu or country-specific top-level domain, used non-standard ports (i.e., not port 80), or pointed to resources with uncommon or deprecated extensions (e.g., .shtml, .ps, .txt).

## 1 Introduction

D-Lib Magazine plays a pivotal role in the documenting and advancing of trends in the digital library community [3]. Given its importance to the community, appropriate measures have been taken to preserve the primary contents of D-Lib Magazine; it is officially mirrored in six other locations throughout the world. D-Lib Magazine is highly interlinked with other digital libraries and the general web. It is published on-line, and all its articles are HTML formatted, thereby making it convenient and attractive for authors to reference web resources by means of hyperlinks. Although the contents of D-Lib Magazine are properly preserved, D-Lib Magazine does not correct external links that become broken over time because of the large effort required to do so. How well do these external links persist over time?

The objective of this paper is to examine the causes of inaccessible links (often referred to as linkrot) contained in D-Lib Magazine articles. We will investigate what causes a link to "go bad" by examining the characteristics of a broken URL. We will examine the URL's age, top-level domain, file name extension, port number, and path characteristics (depth and usage of characters like '~' and '?').

## 2 Related Work

This study is based on a range of previous, related efforts to study the persistence of URLs used in academic online resources. Although not directly related to academic URLs, Koehler [6,7] provides possibly the longest continuous study of URL persistence using the same set of 361 URLs randomly obtained in December 1996. Koehler found a half-life of approximately 2 years. One of the earliest URL persistence studies was performed by Harter and Kim [5]. They examined 47 URLs from scholarly e-journals that were published from 1993 to 1995 and found that one third of the URLs were inaccessible in 1995. Another study [9] monitored 515 URLs that referenced scientific content or education from 2000-2001 and found 16.5% of the URLs became inaccessible or had their content changed. Rumsey examined 3406 URLs used in law review articles published in 2001-1997 and found 52% of the URLs were no longer accessible in 2001 [14]. The persistence of 1000 digital objects (using URLs) from a collection of digital libraries was tested by

Nelson and Allen [11]. Because they were testing URLs that resolved to resources that were "protected" by a digital library, they found only a 3% loss (after manual searching) in URL accessibility during 2000-2001.

Two studies have focused on persistence of URLs that reference computer science and related material. Lawrence et al. [8] performed a study using 67,577 URLs from the computer science articles obtained from the CiteSeer ResearchIndex database [4] that were published from 1993 to 1999. In May 2000 they accessed each of the 67,577 URLs and found the percentage of invalid URLs increased from 23% in 1999 to 53% in 1994. Spinellis [16] performed a similar study using 4224 URLs from computer science articles obtained from the ACM and IEEE Computer Society on-line digital libraries that were published from 1995 to 1999. They performed their experiments once in June 2000 and once again in August 2002. They found the half-life of a referenced URL to be approximately 4 years from the publication date. The study also showed that URLs from 1995-1996 aged at a quicker rate than did URLs from 1997-1998, probably because authors began to cite more persistent URLs and because of improved web site maintenance.

## 3 Methodology

We designed our experiment as follows:

1. From the http://www.dlib.org/back.html page we extracted all the back issues from July 1995 to August 2004, obtaining a total of 101 issues published over the course of approximately 10 years.
2. We downloaded all stories and articles published in the 101 issues, obtaining a total of 453 articles.
3. From these 453 articles we then extracted all the hyperlinks and associated URLs. We removed duplicate URLs per article, producing a total of 7094 URLs, an average of 15.7 per article (median 12, mode 4). The most number of URLs cited by an article was 103, and the least was 1.
4. From the set of 7094 URLs we removed all URLs that referenced www.dlib.org (these were in the form of `http://dx.doi.org/10.1045/*` and `http://www.dlib.org/*`) and all redundant URLs, producing a total of 4387 URLs. There were 77 URLs from doi.org, 1718 URLs from dlib.org, and 912 redundant URLs that were removed.
5. Finally we downloaded each of the 4387 URLs 72 times (three times a week for 25 weeks), beginning on September 9, 2004 and ending on February 27, 2005[1]. We kept track of the http response codes and byte size for each download.

We considered a variety of factors when checking for redundant URLs. We considered `http://foo.edu` and `http://foo.edu/` to be redundant since they both will always resolve to the same location. But the URLs `http://foo.edu/bar` and `http://foo.edu/bar/` were *not* considered to be redundant since it is possible for those URLs to resolve to two different things. For example, both URLs could initially resolve to the directory 'bar'. But if at a later time 'bar' was changed to a text file, the first URL would still resolve, but the second URL would not. We also did not consider `http://foo.edu/` and `http://foo.edu/index.html` to be redundant URLs even though they likely resolve to the same resource. Although the former URL would persist if the foo.edu web server was reconfigured to return index.cgi or default.htm instead of index.html, the latter URL would break.

We decided to exclude all the URLs that referenced www.dlib.org because dlib.org is actively preserved and mirrored and does not reflect natural web conditions. Of the 7094 reference URLs we extracted, 25.3% of them pointed to dlib.org pages, and we found that 98.4% of these URLs were accessible a few days before the first day of testing. We tested the dlib.org URLs again on August 4, 2005, and the same number continued to be accessible. Of those URLs that did not resolve, a manual inspection revealed that 5 of the 17 contained obvious typos that, when corrected, did resolve[2]. Our study focuses on those URLs that point outside of dlib.org.

---

[1] We discarded the results from three trials due to suspected problems with the local machine: 12/17/2004, 12/19/2004, 1/28/2005.
[2] The broken internal dlib.org links were corrected in August 2005 after we notified the editor.

## 4 Results

### 4.1 Scheme Distribution

Table 1 lists the distribution of schemes in the unique set of 4387 URLs. An overwhelming majority of the URLs used the 'http' scheme (99%), whereas a small minority (1%) relied on schemes like 'gopher' and 'ftp' which have fallen out of favor and 'file' which is often used incorrectly.

**Table 1.** URL Schemes

| URL Scheme | Number | Percent |
|---|---|---|
| http | 4326 | 98.61 % |
| ftp | 44 | 1.00 % |
| gopher | 13 | 0.30 % |
| https | 2 | 0.05 % |
| file | 2 | 0.05 % |
| Total | 4387 | 100 % |

### 4.2 Availability at Checkpoints

We found the number of successful downloads varied each time we checked the availability of the 4387 URLs during the 25 week testing period. Figure 1 plots the number of unavailable URLs at each check. There were 1218 (27.8%) unavailable URLs at the first check. By the last check, the number had increased to 1294 (29.5%). There was a sharp and temporary jump on October 10 of 3.4% from the previous test. The jump was primary due to an increase in 500 (internal server error) codes from http://www.ehr.nsf.gov and http://www.nsf.gov which routinely had returned other responses on previous and subsequent checks.

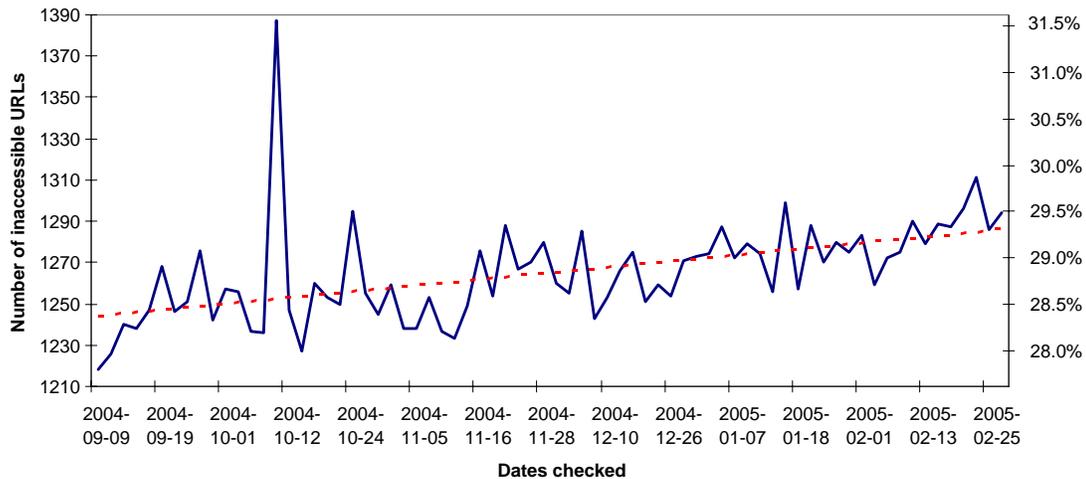

**Fig. 1.** Inaccessible URLs at each check

Although our testing period ran for only 6 months, we were able to observe firsthand how URLs age and become less available over time. The number of URLs that were inaccessible increased by an average of 1.1 at each check as illustrated by the dashed trend line in Fig. 1. As one would expect, there is a strong relationship between a URL's age and its availability. This relationship can be seen more distinctly in Fig. 2 where we plot the accessibility of all the URLs at each checkpoint, grouping the URLs based on the year of publication. A solid point indicates that the URL was accessible. A solid vertical line indicates that a URL was accessible at every check point. Notice that the later years have many more solid vertical lines than the earlier years. Many URLs also fluctuated from accessible to inaccessible and back as indicated by the many "holes" in each vertical line.

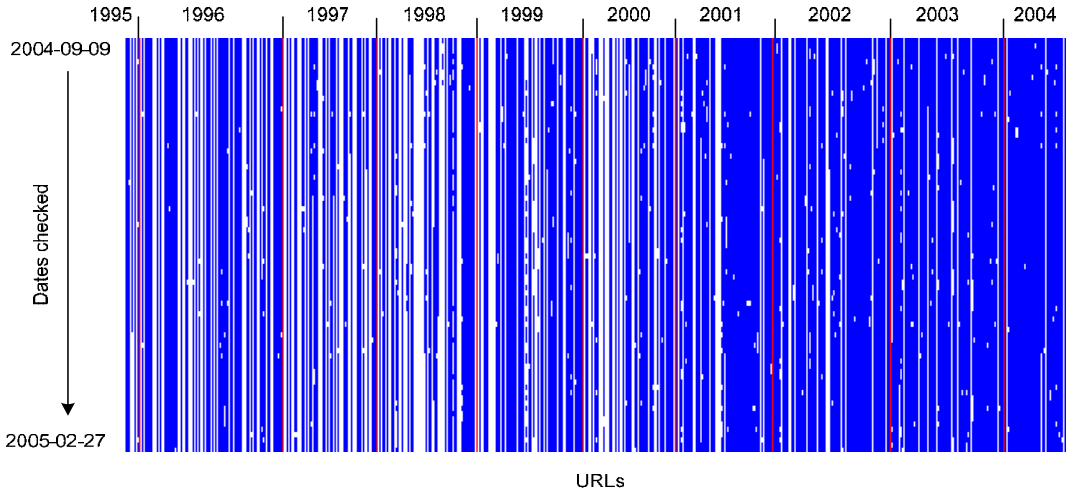

**Fig. 2.** Accessible (solid) and inaccessible (white) URLs at each check by publication year

### 4.3 Distribution by Year

The number of accessible and inaccessible URLs from each publication year is shown in Fig. 3. This figure and all remaining figures in this paper are based on data obtained from our final round of testing and does not include any self-referencing dlib.org URLs. There is a sharp decline in the number of URLs referenced from 1996 to 1997, but most articles cite between 11-15 non-dlib.org URLs per article. The low number of URLs for years 1995 and 2004 are due to the fact that there were only 19 articles (6 months worth) available from 1995 and only 25 articles (8 months worth) available from 2004 when we began testing.

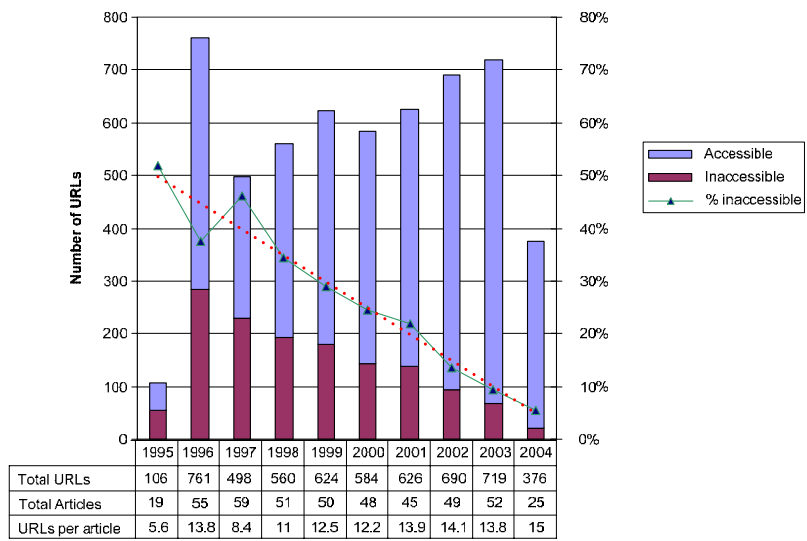

**Fig. 3.** Distribution of URLs by publication year

The number of inaccessible URLs at our last check was only 6% from 2004, but that number jumped to 52% for 1995 URLs. From 2004 to 1995 we see roughly a 5% linear annual increase of inaccessible URLs. As indicated by the dashed trend line, 50% of the URLs are inaccessible 10 years after the publication date. Therefore we report that the half-life of our set of URLs is 10 years from the date of publication. That is, every 10 years after the publication date about half of the URLs will no longer be accessible.

### 4.4 Error Codes

The http status codes of unavailable URLs on the first and last days of testing and are summarized in Table 2. The most common type of status code was 404, meaning that the resource no longer existed on the server. The second most common cause was that the server encountered an unexpected condition that prevented it from fulfilling the request, indicated by error code 500. The error codes 404 and 500 together made up about 95% of the failures. Other failures we encountered in very small numbers were 400 (bad request) and 503 (service unavailable).

**Table 2.** Composition of returned error codes

| HTTP Code | Meaning | 2004-09-09 | 2005-02-27 |
|---|---|---|---|
| 404 | Not found | 62.40 % | 60.20 % |
| 500 | Internal sever error | 32.51% | 35.09 % |
| 403 | Forbidden | 3.94 % | 3.86 % |
| 401 | Unauthorized | 0.74 % | 0.62 % |
| 200 | OK but 0 length content | 0.25 % | 0.23 % |
| 410 | Gone | 0.08 % | 0.00 % |
| 502 | Bad gateway | 0.08 % | 0.00 % |

The http code 200 was included in the error codes because occasionally a web server will return a zero length byte stream with an http 200 code. We categorize URLs with no content as inaccessible. There was one URL (ftp://ftp.math.psu.edu/pub/sib) that consistently returned a 200 code, but its size fluctuated from 0 to a non-zero value. We later discovered that it was because the server had a limit on the maximum number of simultaneous connections. We did not test for "soft 404s", pages that returned a 200 but indicated the resource was not found [1].

### 4.5 Path Depth

To determine how a URL's path length influences decay rates, we calculated the path depth for each accessible and inaccessible URL. To calculate path depth, we added 1 to the depth for every directory or file after the domain name. For example, http://foo.com/ has a path depth of 0, http://foo.com/bar.html has a depth of 1, http://foo.com/dir/bar.html has a depth of 2, etc. We also added 1 to the path depth for any existing query string in a URL (e.g., http://foo.org/cgi?bar=2 has a path depth of 2).

Figure 4 shows the distribution of the 4387 URLs according to path depth. A majority (66%) of the URLs have a path depth of 2 or less.

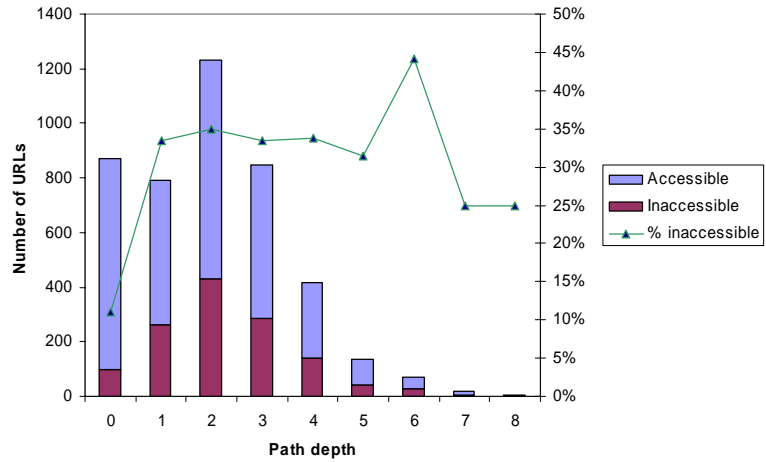

**Fig. 4.** Distribution of URLs by path depth

Figure 4 shows a significant increase (22%) in inaccessible URLs as the path depth moves from 0 to 1. An inaccessible URL with path depth 0 implies the disappearance of an organization or business (either it has changed its name or gone out of business). An inaccessible URL with path depth greater than 0 would more likely be caused by the reorganization of internal structure or system change. The persistence of deep hierarchical URLs (depth 7 and 8) can possibly be explained by the extra attention and organization required by a web site administrator to maintain the URLs. The spike at path depth 6 is due to the inaccessibility of the web server cs-tr.cs.cornell.edu which made up 37% of the inaccessible URLs, and the use of non-standard ports which made up an additional 23% of the inaccessible URLs.

### 4.6 Top-Level Domain

The distribution of URLs by top-level domain name is shown in Fig. 5. A majority of URLs referenced items in the .edu domain (28%) and .org domain (25%). This is likely because most scholarly material is available at these top-level domains. URLs using county code domains were not frequently used, and several of them (e.g., .au, .nl, .ca, .us) were 35% or more inaccessible. It is not surprising that the .edu URLs had the fourth highest percentage of broken URLs (30%, behind .nl, .au. and .net) due to the transient nature of university life where professors, students, and research projects are frequently coming and going. Our findings are consistent with [14] and [9,10] who also found .org URLs persisting longer than .edu URLs.

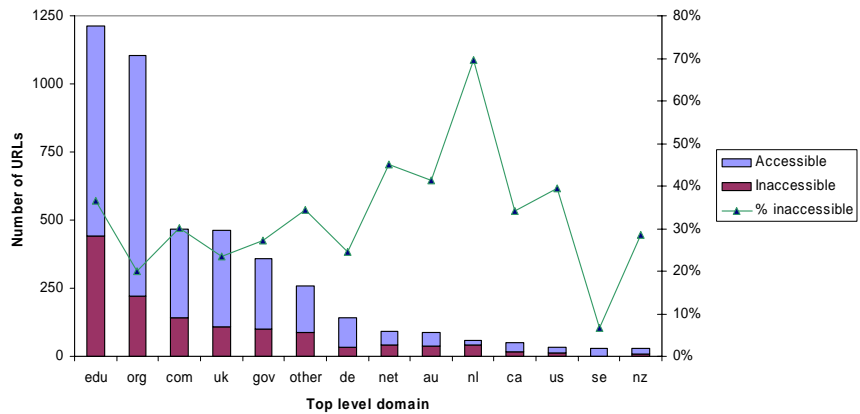

**Fig. 5.** Distribution of URLs by top-level domain

### 4.7 Path Characteristics

We examined the following characteristics of each URL to determine their effects on persistence:

1. Existence of a '~' after the domain name that indicates a personal homepage (e.g., http://www.foo.edu/~homepage).
2. Existence of a ':number' that indicates use of a non-standard port (e.g., http://www.foo.edu:8080).
3. Existence of a '?' that indicates use of a query string to produce dynamic web pages (e.g., http://www.foo.edu?id=123)

The results are shown in Table 3. The result of a chi-square analysis indicate a statistically significant relationship between all 3 factors and the inaccessibility and accessibility of web pages ($p<0.001$). URLs with non-standard ports (83%) were the most likely to disappear over time. Homepage URLs faired much better but were still found to be inaccessible almost half of the time. This is probably due to the continual shifting of personnel and students from one organization to another. Finally, dynamically produced web pages proved to be the most durable but still were not as persistent as those that appear to be static (29% inaccessible).

**Table 3.** Distribution of URLs referencing personal home pages, non-standard ports, and dynamic pages

|  | Personal home page | Non-standard port | Dynamic page |
|---|---|---|---|
| Inaccessible URLs | 136 | 53 | 76 |
| Accessible URLs | 126 | 11 | 109 |
| Total URLs | 262 | 64 | 185 |
| % Inaccessible | 51.9 % | 82.8 % | 41.1 % |

### 4.8 File Extension

The distribution of URLs by file extension can be seen in Fig. 6. URLs that use query strings (i.e., contain the '?' character) are grouped with 'other'. By far the two most popular file extensions are the slash (i.e., URLs that end with a slash as in http://foo.edu/) and .htm or .html. These account for 76% of all the URL file extensions used in the collection of 4387 URLs.

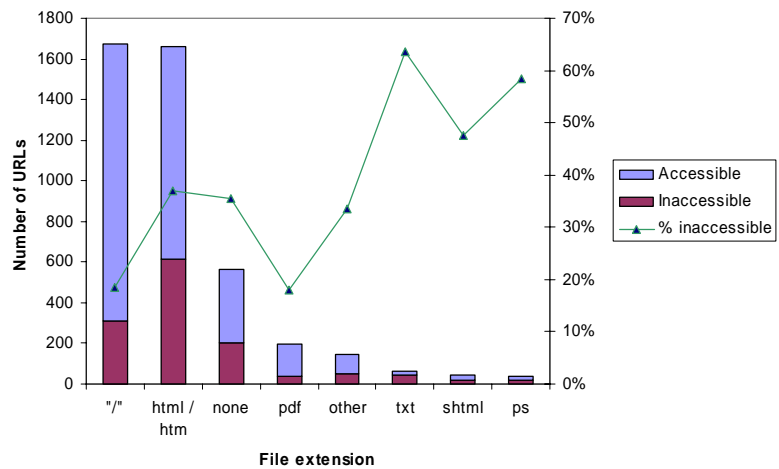

**Fig. 6.** Distribution of URLs by file name extension

The use of '/' and .pdf showed to be the most persistent. The success of '/' is probably due to the fact that most directories either have a default file that is served (e.g., index.html, default.htm) or a listing of the directory's content. The success of .pdf URLs is due to its newness. PDF has become the de facto standard for storing scholarly material in a system independent manner, and only recently have .pdf URLs started being used. Only one .pdf URL was used in 1996, and 7 in 1998. In 2004, 58 .pdf URLs were used, an average of 2 per article.

The highest inaccessibility levels were observed for URLs with file extensions .txt, .shtml and .ps. The .txt extension is typically used for temporary textual data that may have been converted into a web page later on or simply disregarded as it became stale. The .shtml and .ps extensions are indicative of file formats that are slowly becoming obsolete.

### 4.9 Use of Persistent URLs

To provide more stable URLs, on-line resource providers have begun using Persistent URLs (PURLs) [15], handles [17], and Digital Object Identifiers (DOIs) [13] to allow a resource to change locations without changing its URL. Unfortunately there were few PURLs, handles, and DOIs in the collection of D-Lib URLs. There were only 89 PURLs in the complete set of 7094 URLs from all articles, and only 57% of these were accessible. There were only 2 handles (both were accessible), and there were only 15 DOIs that did not point back to dlib.org (all were accessible).

### 4.10 Content Changes

We monitored the size of the content returned from each URL to determine its stability. During the 25 week testing period, 1520 of the 4387 URLs (34.6%) we were monitoring registered at least one size change (only successful downloads were included). An average of 334 of the 1520 URLs (22.0%) produced different byte sizes at each check. Of these 1520 URLs, 683 of them (44.9%) registered a change of at least 1 KB. We will define a URL to be of the "in flux" category if at each check its non-zero size had changed from the last observation. Some "in flux" URLs changed sizes constantly; 51 of the "in flux" URLs returned a different size at *every* check.

Figure 7 shows the cumulative effect of content size changes at each URL checkpoint for the 683 "in flux" URLs. The large increase at check number 11 and subsequent decrease at check number 30 were largely due to a single URL (http://www.cs.cornell.edu/cdlrg/Reference%20Linking/tr1842.ps) that fluctuated by 4 MB. We performed a manual inspection of 10% of the 683 URLs that fell into the "in flux" list category and found that the primary reason for content size fluctuations was due to dynamically generated advertisements. Also some web sites frequently changed the images they were displaying which contributed to the fluctuations.

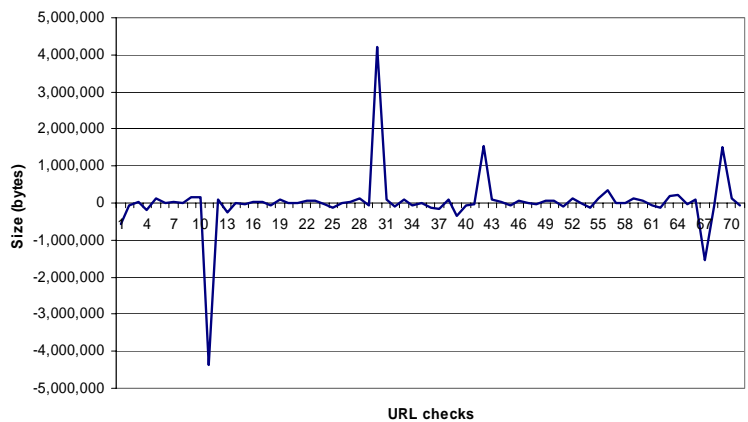

**Fig. 7.** Cumulative content size fluctuation of all URLs

# 5 Discussion

Using the publication date of inaccessible URLs, we found the half-life of a URL web reference from D-Lib Magazine to be 10 years. This result shows better URL persistence compared to the 4 year half-life reported by Spinellis [16]. The longer half-life indicates that the D-Lib Magazine URLs are persisting longer than the sample URLs from articles in IEEE Computer and Communications of the ACM. Possibly the authors of D-Lib Magazine articles may have taken more care when choosing to use URLs because digital preservation and persistence is frequently a concern for digital library practitioners. Another factor to consider is that possibly greater care is currently being taken by web site administrators to create more persistent URLs. Many of our URLs were from a later time period (1995-2004) than were the URLs from the Spinellis study (1995-1999).

Our findings also differ from the Spinellis study when comparing path depth. Spinellis found that as URL path depth increased, so did the failure rate for his set of URLs. Our data shows a similar effect, but only for path depths up to 2. URLs with path depths greater than 2 tend to level off and not experience a significant amount of increased unavailability.

Our findings also showed that the D-Lib URLs had a longer half-life than the CiteSeer URLs reported by Lawrence et al. [8]. They found a high percentage of accessible URLs in 1993 and explained that this was likely due to the fact that the web was new, and many citations were only to well-known sites like http://www.intel.com/. Ignoring the 1993 data, we calculated the half-life for a computer science article URL in the Lawrence study to be approximately 6 years from the publication date. We calculate a half-life of about 17 years if the 1993 data is included. The Lawrence study considered any string that began with (http:|https:|ftp:) to be a URL which introduced an unknown number of unintended URLs into their dataset. For example, we have used numerous URLs like http://foo.edu in this paper but did not intend for them to resolve to a specific resource. When 300 invalid URLs were randomly sampled from their collection, they found 32% contained syntax errors or were example URLs like http://foo.edu. (We only found a small handful of syntactically invalid URLs in our experiment since we extracted directly from hyperlinks.) This means the approximate 6 year half-life is probably lower than it should be, but is still not near the 10 year URL half-life we are reporting.

Both the Spinellis and Lawrence studies showed significantly higher initial decay rates than what we measured. The Spinellis study showed a 20% initial decay and then 10% decay per year for the next three years. The Lawrence study showed a 23% initial decay and then 5% decay per year for 5 years. We initially experienced 6% decay followed by 5% decay per year. Our initial decay is possibly lower because the lag time between writing an article and publishing it is lower for D-Lib Magazine than for other publications.

It has been well established that the web is extremely ephemeral. A recent study [12] concluded that only about 20% of today's web pages will be accessible one year from now. Considering how frequently web pages disappear, authors should be very careful what URL citations they use in their work. Lawrence et al. [8] recommends providing formal citations along with URL citations when possible and providing context information to improve web searches for missing resources. Authors making their work available are suggested to place resources in a central repository and avoid URLs that depend on a personal directory or machine name. Spinellis [16] recommends the creation of recommendations from publishers and professional societies to provide guidance on using web citations properly. URL verification should be given a higher priority in the article editing process, and authors should cite published versions of a work instead of online versions and cite material in organized repositories instead of material on personal web sites. Spinellis additionally suggests the standardization of HTML meta-tags so web administrators could indicate the projected persistence of online material. Berners-Lee offers some advice to web administrators on how to better organize their URIs [2].

Our study revealed several general characteristics that are indicators of unstable URLs. The URL characteristics below were associated with increased levels of linkrot:
- a non-standard port
- a personal homepage
- dynamic query strings
- uncommon or deprecated file extensions (e.g., .txt, .shtml, .ps)
- .net, .edu or country-specific top-level domain names

## 6  Conclusions

About 28% of the 4387 unique URL references from the D-Lib Magazine articles between 1995 and 2004 were inaccessible on 2004-9-9, and that number grew to 30% on 2005-2-27. The most common cause of failure was "404 not found", followed by the "500 internal server error". In addition, the association between the disappearance of a URL and its age was strong. Based on the collected data, the half-life of a D-Lib Magazine article URL (excluding URLs that point back to dlib.org) is 10 years. This is longer than the previously reported values of 4 years for articles from Computer and CACM [16] and the 6 years for articles from CiteSeer we calculated from [8]. Based on the collected data, pages in the .net, .edu and other country-specific top-level domains were more likely to disappear than all other top-level domains. In addition, internal structural or system change was more likely to be the reason for URL disappearance since the percentage of URL disappearance increases sharply from the path depth of 0 to 1. Also the likelihood of URL disappearance increases until the path depth of 2 and then start to decrease. This indicates the extra effort spent on organization may out-weigh the accumulation of each element of the URL path's probability to fail. Furthermore, URLs with .html file name extension or '/' at the end were more likely to persist than other URLs. URLs that use non-standard ports, link to personal homepages, or produce dynamic content are more likely to disappear.  Finally about 84% of all URLs had size fluctuations that were less than 1 KB, and the inclusion of advertisement banners accounted for the majority of content size fluctuations.

## 7  Acknowledgements

We would like to thank Bonnie Wilson, editor of D-Lib Magazine, for reviewing our paper.